\documentclass[11pt]{article}
\usepackage{moriond,epsfig}

\def\Journal#1#2#3#4{{#1} {\bf #2}, #3 (#4)}

\def\NIM{\em Nucl. Instrum. Methods}

\def\PLB{{\em Phys. Lett.}  B}
\def\PRL{\em Phys. Rev. Lett.}
\def\PRD{{\em Phys. Rev.} D}

\def\be{\begin{equation}}
\def\ee{\end{equation}}
\def\bea{\begin{eqnarray}}
\def\eea{\end{eqnarray}}

\begin{document}
\vspace*{4cm}
\title{SOLAR NEUTRINO PHYSICS WITH BOREXINO I}

\author{L.\ LUDHOVA$^{1}$ , G.~BELLINI$^{1}$, J.~BENZIGER$^{2}$,  D.~BICK$^{3}$, G.~BONFINI$^{4}$, D.~BRAVO$^{5}$, M.~BUIZZA~AVANZINI$^{1}$, B.~CACCIANIGA$^{1}$,  L.~CADONATI$^{6}$, F.~CALAPRICE$^{7}$, C.~CARRARO$^{8}$,  P.~CAVALCANTE$^{4}$,  A.~CHAVARRIA$^{7}$, D.~D{\textquoteright}ANGELO$^{1}$, S.~DAVINI$^{8,9}$, A.~DERBIN$^{10}$, A.~ETENKO$^{11}$, K.~FOMENKO$^{4,12}$, D.~FRANCO$^{13}$, C.~GALBIATI$^{7}$, S.~GAZZANA$^{4}$, C.~GHIANO$^{4}$, M.~GIAMMARCHI$^{1}$, M.~G\"{O}GER-NEFF$^{14}$,  A.~GORETTI$^{7}$,  L.~GRANDI$^{7}$, E.~GUARDINCERRI$^{8}$, S.~HARDY$^{5}$, ALDO~IANNI$^{4}$, ANDREA~IANNI$^{7}$, A.~KAYUNOV$^{10}$,V.~KOBYCHEV$^{17}$,  D.~KORABLEV$^{12}$,  G.~KORGA$^{9}$,   Y.~KOSHIO$^{4}$, D.~KRYN$^{13}$, M.~LAUBENSTEIN$^{4}$,  T.~LEWKE$^{14}$, E.~LITVINOVICH$^{11}$,  B.~LOER$^{7}$, F.~LOMBARDI$^{4}$,  P.~LOMBARDI$^{1}$,  I.~MACHULIN$^{11}$, S.~MANECKI$^{5}$, W.~MANESCHG$^{15}$, G.~MANUZIO$^{8}$, Q.~MEINDL$^{14}$,  E.~MERONI$^{1}$, L.~MIRAMONTI$^{1}$, M.~MISIASZEK$^{16,4}$, D.~MONTANARI$^{7,4}$,  P.~MOSTEIRO$^{7}$, V.~MURATOVA$^{10}$,  L.~OBERAUER$^{14}$,  M.~OBOLENKSY$^{13}$, F.~ORTICA$^{18}$,  K.~OTIS$^{6}$,  M.~PALLAVICINI$^{8}$,  L.~PAPP$^{4,5}$,    L.~PERASSO$^{8}$, S.~PERASSO$^{8}$, A.~POCAR$^{6}$, R.S.~RAGHAVAN$^{\dag 5}$\footnote{$^{\dag}$ deceased}~, G.~RANUCCI$^{1}$, A.~RAZETO$^{4}$,  A.~RE$^{1}$, P.A.~ROMANI$^{18}$, A.~SABELNIKOV$^{11}$,  R.~SALDANHA$^{7}$,  C.~SALVO$^{8}$,   S.~SCH\"ONERT$^{14}$, H.~SIMGEN$^{15}$, M.~SKOROKHVATOV$^{11}$, O.~SMIRNOV$^{12}$, A.~SOTNIKOV$^{12}$, S.~SUKHOTIN$^{11}$,  Y.~SUVOROV$^{4}$,  R.~TARTAGLIA$^{4}$,  G.~TESTERA$^{8}$, D.~VIGNAUD$^{13}$,  R.B.~VOGELAAR$^{5}$,  F.~VON~FEILITZSCH$^{14}$, J.~WINTER$^{14}$, M.~WOJCIK$^{16}$,  A.~WRIGHT$^{7}$, M.~WURM$^{3}$, J.~XU$^{7}$, O.~ZAIMIDOROGA$^{12}$,  S.~ZAVATARELLI$^{8}$, AND  G.~ZUZEL$^{16}$}

\address{$^{1}$Dipartimento di Fisica, Universit\'{a} degli Studi e INFN, Milano 20133, Italy\\
$^{2}$Chemical Engineering Department, Princeton University, Princeton, NJ 08544, USA\\
$^{3}$Institut f\"ur Experimentalphysik, Universit\"at Hamburg, Germany\\
$^{4}$INFN Laboratori Nazionali del Gran Sasso, Assergi 67010, Italy\\
$^{5}$Physics Department, Virginia Polytechnic Institute and State University, Blacksburg, VA 24061, USA\\
$^{6}$Physics Department, University of Massachusetts, Amherst 01003, USA\\
$^{7}$Physics Department, Princeton University, Princeton, NJ 08544, USA\\
$^{8}$Dipartimento di Fisica, Universit\'{a} e INFN, Genova 16146, Italy\\
$^{9}$Department of Physics, University of Houston, Houston, TX 77204, USA\\
$^{10}$St. Petersburg Nuclear Physics Institute, Gatchina 188350, Russia\\
$^{11}$NRC Kurchatov Institute, Moscow 123182, Russia\\
$^{12}$Joint Institute for Nuclear Research, Dubna 141980, Russia\\
$^{13}$APC, Laboratoire AstroParticule et Cosmologie, 75231 Paris cedex 13, France\\
$^{14}$Physik Department, Technische Universit\"{a}t M\"{u}nchen, Garching 85747, Germany\\
$^{15}$Max-Plank-Institut f\"{u}r Kernphysik, Heidelberg 69029, Germany\\
$^{16}$M. Smoluchowski Institute of Physics, Jagiellonian University, Krakow, 30059, Poland\\
$^{17}$Kiev Institute for Nuclear Research, Kiev 06380, Ukraine\\
$^{18}$Dipartimento di Chimica, Universit\'{a} e INFN, Perugia 06123, Italy}

\maketitle\abstracts{
Borexino is a large-volume liquid scintillator detector installed in the underground halls of the Laboratori Nazionali del Gran Sasso in
Italy. After several years of construction, data taking started in May 2007. The Borexino phase I ended after about three years of data taking. Borexino provided the first real time measurement of the $^{7}$Be solar neutrino interaction rate with accuracy better than 5\% and confirmed the absence of its day--night asymmetry with 1.4\% precision. This latter Borexino results alone rejects the LOW region of solar neutrino oscillation parameters at more than 8.5 $\sigma$ C.L. Combined with the other solar neutrino data, Borexino measurements isolate the MSW--LMA solution of neutrino oscillations without assuming CPT invariance in the neutrino sector. Borexino has also directly observed solar neutrinos in the 1.0-1.5 MeV energy range, leading to the first direct evidence of the $pep$ solar neutrino signal and the strongest constraint of the CNO solar neutrino flux up to date. Borexino provided the measurement of the solar $^{8}$B neutrino rate with 3 MeV energy threshold. 
}

\section{Introduction}

Solar neutrinos ($\nu_{e}$) are expected to be produced in the two distinct fusion processes, in the main $pp$ fusion chain and in the sub-dominant CNO cycle. In the past 40 years, solar neutrino experiments~\cite{solar} have revealed important information about the Sun~\cite{ssm,ssm_new} and have shown that solar $\nu_{e}$ undergo flavor transitions that are well described by Mikheyev--Smirnov--Wolfenstein Large Mixing Angle (``MSW--LMA'') type flavor oscillations~\cite{msw}. Reactor anti--neutrino ($\bar{\nu}_{e}$) measurements~\cite{kamland} also support this model.  The MSW model predicts a transition in the solar $\nu_e$ survival probability $P_{ee}$ at energies of about 1-4\,MeV from vacuum--dominated to matter--enhanced oscillations. This transition is currently poorly tested. Therefore, in order to test MSW-LMA thoroughly, to probe other proposed ${\nu}_{e}$ oscillation scenarios~\cite{nonstandard}, and to further improve our understanding of the Sun (metallicity problem~\cite{smp,ssm_new}), it is important to measure the solar ${\nu}_{e}$ fluxes.

Borexino is a real--time large--volume  liquid scintillator detector~\cite{detpa} installed in the underground halls of Laboratori Nazionali del Gran Sasso in Italy (3800 m water equivalent). It was designed to measure the 862~keV $^7$Be solar $\nu_{e}$. One of its unique features is the very low background level that allowed the first $^7$Be-$\nu_e$ measurement~\cite{be7_1} soon after the detector became operational in May 2007. This made Borexino the first experiment capable of making spectrally resolved measurements of solar $\nu_{e}$'s at energies below 1\,MeV. Borexino performed also a measurement of the $^8$B solar $\nu_{e}$'s with a recoil--electron energy threshold of 3 MeV~\cite{b8}. Recent Borexino solar $\nu_e$ results include a high--precision measurements of the $^7$Be--$\nu_{e}$ interaction rate~\cite{be7} and of the absence of its day--night asymmetry~\cite{adn}, the first measurement of $pep$-$\nu_{e}$'s and the strongest constraint up to date on CNO solar neutrinos~\cite{pep}. Other Borexino results include the study of solar and other unknown $\bar{\nu}_{e}$ fluxes~\cite{antinu}, observation of geo--neutrinos~\cite{geonu}, experimental limits on the Pauli--forbidden transitions in $^{12}$C nuclei \cite{pauli}, and a search for 5.5~MeV solar axions produced in $p(d,^{3}He)A$ reaction~\cite{axions}.

\section{Borexino detector}

Borexino detects low--energy solar $\nu_{e}$ via their elastic scattering off the electrons of a $\sim$280~tons  ultra--pure liquid scintillator, while $\bar{\nu}_{e}$ are detected via the inverse neutron $\beta$--decay reaction, with 1.806~MeV kinematic threshold. The high light yield and the extreme radiopurity achieved allow the real--time detection of solar $\nu_{e}$'s down to about 20~keV of electron recoil energy, being limited below this value by the presence of the unavoidable $^{14}$C($e^-$) background.
 
\begin{figure}[t]
\centering
$
\begin{array}{cc}
\includegraphics[width=0.45\linewidth] {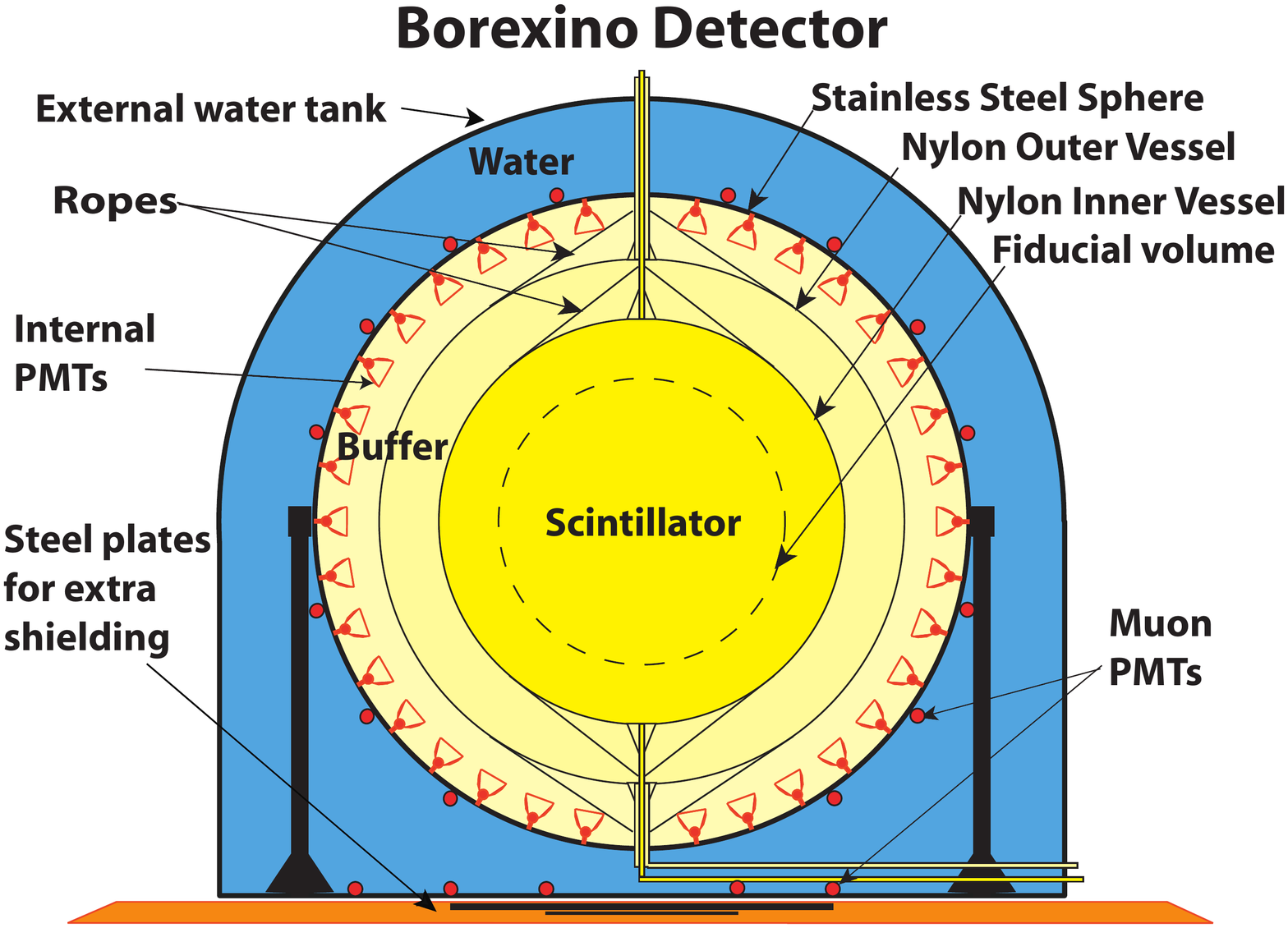}& 
\includegraphics[width=0.55\linewidth, height=.32\linewidth, keepaspectratio=true]{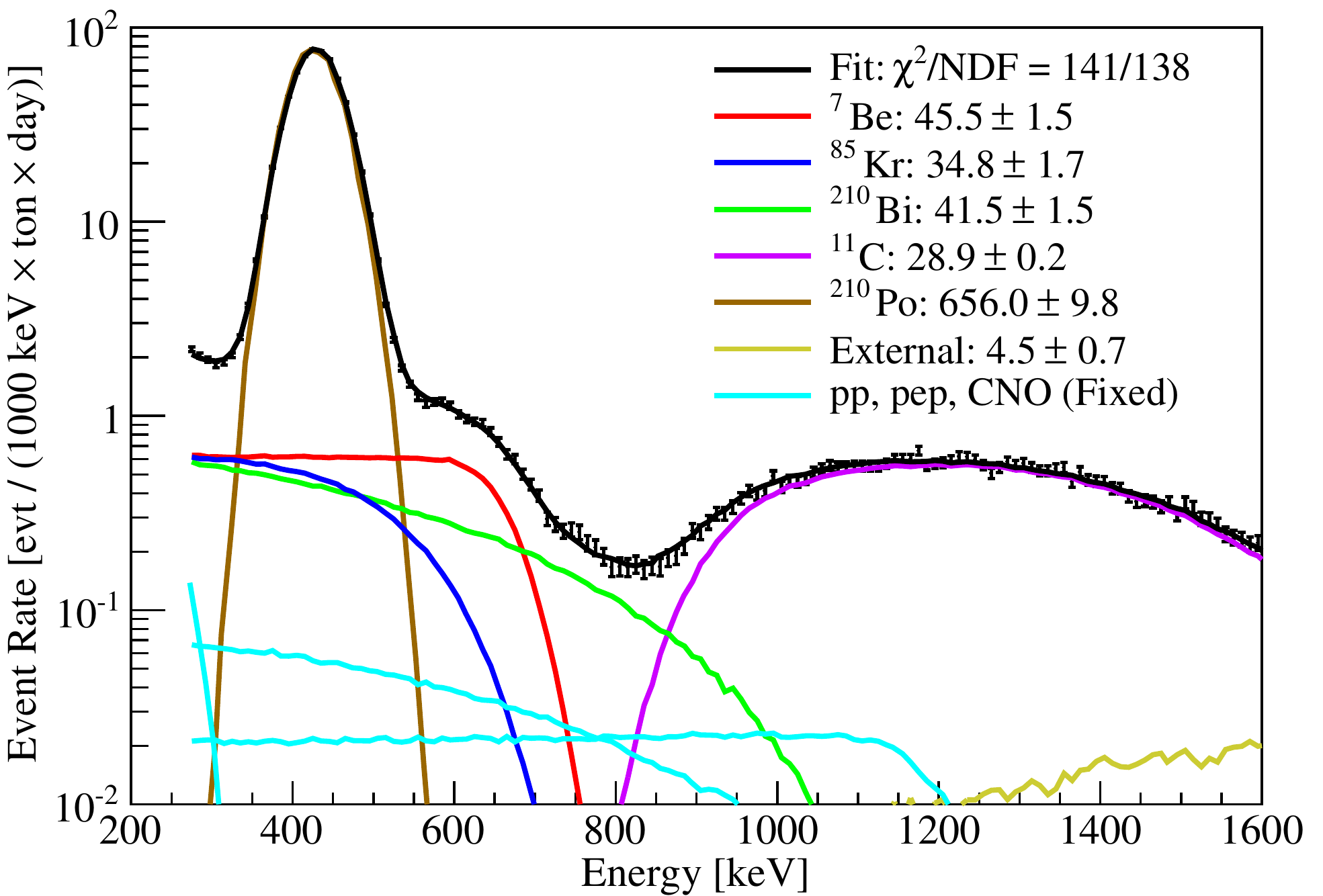}\\
\end{array}
$
\caption{\textbf{Left}: The Borexino detector.
\textbf{Right}: A Monte Carlo based fit over the energy region (270--1600)~keV;
Rate values in the legend are integrated over all energies and are quoted in units of counts/(day$\cdot$100~ton).
}\label{fig:be7}
\end{figure}

The main features of the Borexino detector~\cite{detpa} are illustrated in Figure \ref{fig:be7}-Left. The active medium is a mixture of pseudocumene (PC, 1,2,4- trimethylbenzene) and a wavelength shifter PPO (2,5-diphenyloxazole, a fluorescent dye) at a concentration of 1.5~g/l. The scintillator is contained in a 125~$\mu$m thick nylon vessel with a radius of 4.25~m, shielded by the two PC buffers separated by a second nylon vessel which acts as a barrier against the inward radon diffusion. The 1000 tons of PC buffer contain 5.0 g$/$l DMP (dimethylphthalate) quenching the residual PC scintillation. The scintillator and buffers are contained within a $13.7$~m diameter stainless steel sphere that is housed in a $16.9$~m high cylindrical dome filled with ultra--pure water that serves as an additional passive shielding and as an active Cherenkov muon veto system equipped with 200 PMTs \cite{mupa}. The  scintillation light is viewed by 2212 8$''$ PMTs (ETL9351) mounted on on the inside  surface of the stainless steel sphere. The number of  hit PMTs is a measure of the deposited energy. The position of the scintillation event is determined by a photon time--of--flight method. There is no sensitivity to the intrinsic neutrino direction.

With the muon flux and external background highly suppressed, the crucial requirement for solar $\nu_e$ detection is a high scintillator radiopurity achieved via a combination of distillation, water extraction, and nitrogen gas stripping \cite{purif,detpa}. Assuming secular equilibrium in the Uranium and Thorium decay chains, the Bi--Po delayed coincidence rates imply  $^{238}$U and $^{232}$Th levels of (1.6 $\pm$ 0.1)$\times 10^{-17}$ g/g and (6.8 $\pm$ 1.5)$\times 10^{-18}$ g/g~\cite {be7_2,detpa}. The radon progenies $^{210}$Po and $^{210}$Bi  however, are higher than expected and are out of secular equilibrium. The $^{85}$Kr is present in the data due to a small air leak during the detector filling. Nevertheless, the current low background has made possible measurements of $^7$Be neutrinos with high accuracy, measurements of $^8$B and pep neutrinos, and strong limits on CNO neutrinos. Systematic errors were reduced thanks to extensive calibration campaigns performed deploying $\alpha$, $\beta$, $\gamma$, and neutron sources within the scintillator volume. The calibration data were also used to validate and improve both the Monte Carlo (MC) and the analytic detector response function.

\section{Precision measurement of $^7$Be neutrinos}

The observed energy spectrum based on 740.7 days of life time after cuts is shown in Figure~\ref{fig:be7}-Right.
The apparent shoulder at 665 keV is due to the Compton--like spectrum of recoil $e^-$'s scattered by the 862 keV mono-energetic $^7$Be-$\nu_e$. The peak at  440 keV is due to $^{210}$Po $\alpha$'s. The $^{11}$C  produced by muon interactions on $^{12}$C has the continuous $e^+$ spectrum above 800 keV. The $^7$Be-$\nu_e$ rate was determined by fitting the energy spectrum to the expected $\nu_e$ and background spectra. Two independent methods were used, one MC based and one using an analytic detector response function. In both methods, the weights for the $^7$Be$-\nu_e$ and the main background components ($^{85}$Kr, $^{210}$Po,$^{210}$Bi, and $^{11}$C) were free fit parameters, while those of the $pp$, $pep$, CNO, and $^8$B $\nu_e$'s were fixed to the SSM predictions assuming MSW--LMA oscillations. The MC method includes also external $\gamma$-ray background. The energy scale and resolution were floated in the analytic fits, while the MC approach automatically incorporates the detector response. The stability of the results was studied by repeating the fits with slightly varied characteristics. One example of the MC--fit spectrum is shown in Figure  \ref{fig:be7}-Right.

The best estimate for the $^7$Be--$\nu_e$ interaction rate in Borexino is $(46.0 \pm 1.5 \mbox{(stat)} ^{+1.5}_{-1.6} \mbox{(syst)})$
counts/(day$\cdot$100ton) \cite{be7}; 100 tons of Borexino scintillator contain 3.757 $\times$ 10$^{31}$~$e^-$. The measured rate is to be compared to the predicted rate without $\nu_e$ oscillations of $(74.0 \pm 5.2)$ counts/(day$\cdot$100ton), based on the high metallicity SSM flux~\cite{ssm_new}. The experimental result is 5$\sigma$ lower. The ratio of the measured to the predicted $\nu_e$-equivalent flux is $0.62\pm0.05$. Under the assumption that the reduction in the apparent flux is the result of $\nu_e$ oscillation to $\nu_\mu$ or $\nu_\tau$, we find $P_{ee} = 0.51 \pm 0.07$ at 862~keV. Alternatively, by assuming MSW-LMA solar neutrino oscillations, the Borexino results can be used to measure the $^7$Be solar neutrino flux, corresponding to $\Phi_{^7\mbox{Be}} = (4.84 \pm 0.24) \times 10^9 \mbox{ cm}^{-2} \mbox{ s}^{-1}$.

\section{Search for a day-night asymmetry in the 862 keV $^7$Be neutrino rate}

Borexino data can be used to search for a change in the $^7$Be--$\nu_e$ rate associated with matter effects possibly causing $\nu_e$ regeneration as they pass through the Earth during the night. The asymmetry between the night and day rates, $R_N$ and $R_D$, is described by $A_{dn}$ parameter
\begin{equation}
A_{dn} = 2 \frac{R_N - R_D}{R_N + R_D}
\end{equation}
The result is $A_{dn} = (0.001 \pm 0.012 \mbox{(stat)} \pm 0.007 \mbox{(syst)})$,
fully consistent with zero \cite{adn}. The $\Delta m^2_{12}$ region between $\sim$$10^{-8}$ and $2 \times 10^{-6}$~eV$^2$ is excluded by this result alone. In particular, the minimum $A_{dn}$ computed in the LOW $\Delta m^2$ region is 0.117, more than 8.5$\sigma$ away from our measurement.  The inclusion of this result in a global analysis of all solar $\nu_e$ data can single out the LMA solution of solar $\nu_e$ oscillations with very high confidence. The result does not use the KamLAND $\bar{\nu}_e$ data \cite{kl}, and therefore does not assume CPT invariance in the neutrino sector.

\begin{figure}[t]
\centering
$
\begin{array}{cc}
\includegraphics[width=0.48\linewidth] {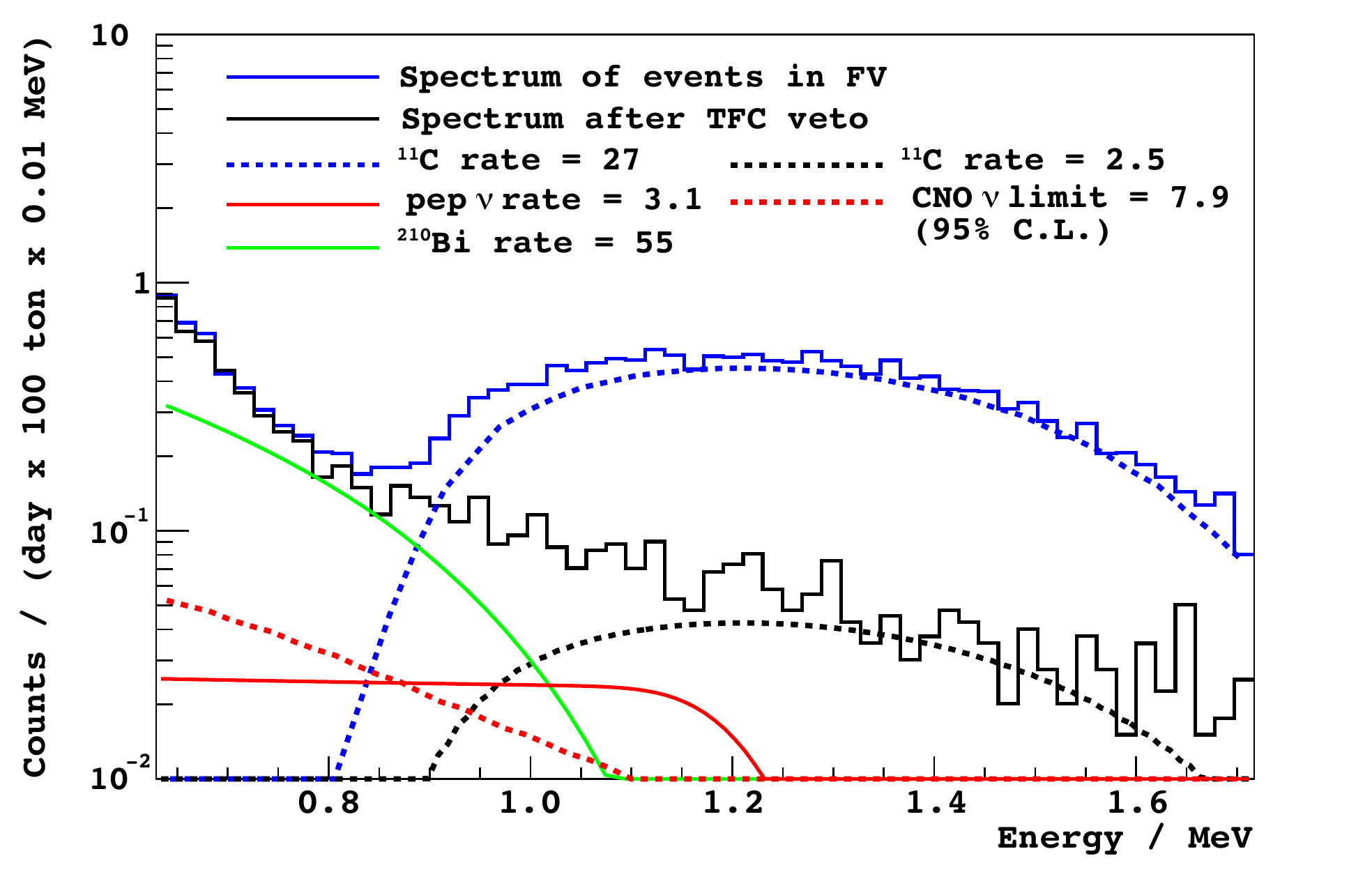}& 
\includegraphics[width=0.52\linewidth, height=.32\linewidth, keepaspectratio=true]{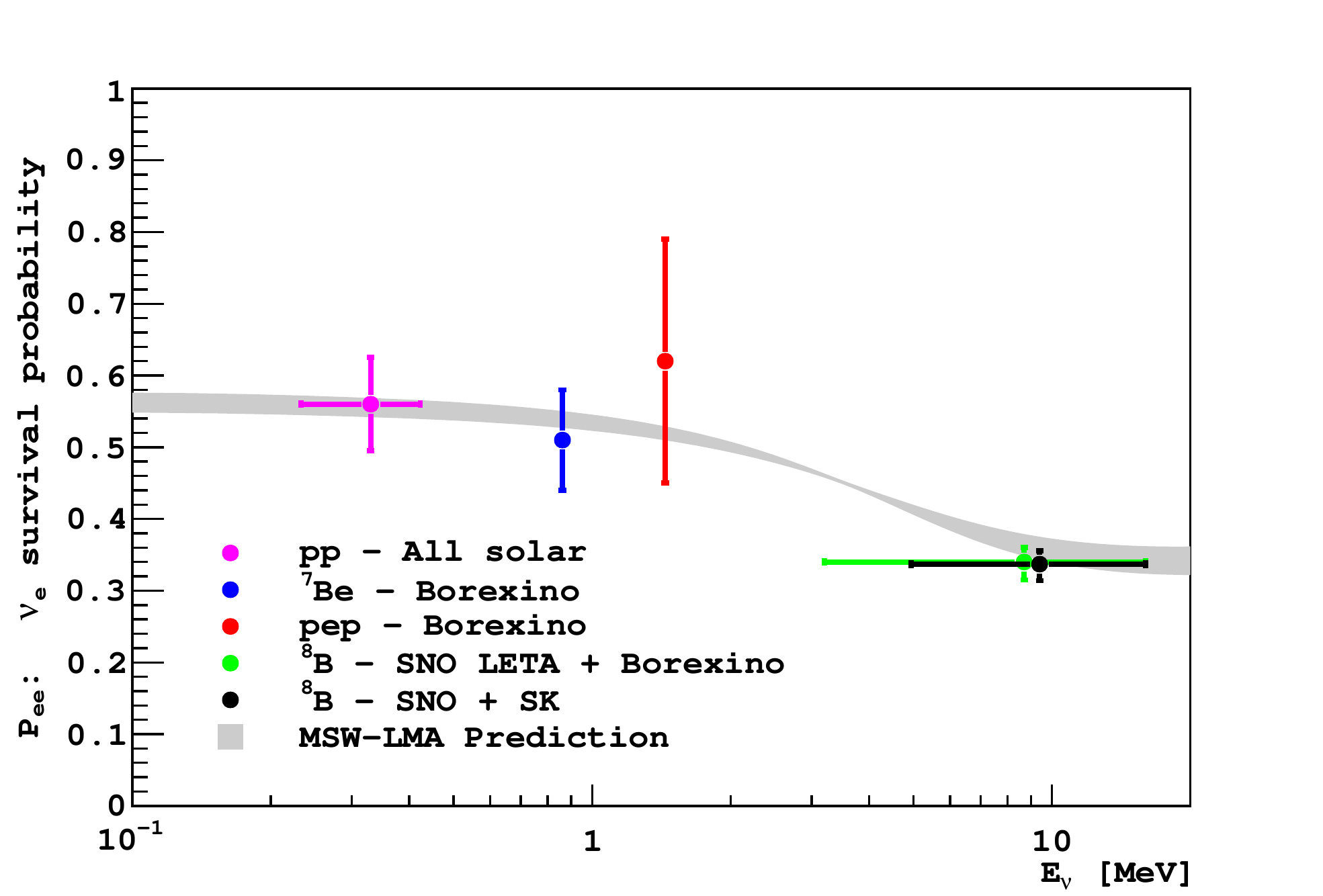}\\
\end{array}
$
\caption{\textbf{Left}: Energy spectra before (solid blue) and after (solid black) the TFC veto. The estimated $^{11}$C rate is shown before (dashed blue) and after (dashed black) the veto. The green line shows $^{210}$Bi. The red lines represent the $pep$--$\nu_e$ best estimate (solid) and the CNO--$\nu_e$  upper limit (dashed). Rates in the legend are integrated over all energies and in units of counts/(day$\cdot$100ton). \textbf{Right}: The $P_{ee}$ as a function of $\nu_e$ energy. The grey band shows the MSW--LMA prediction with 1$\sigma$ range of mixing parameters. The data points are described in the legend. }
\label{fig:pep}
\end{figure}

\section{First evidence of $pep$ solar neutrinos and limit on CNO solar neutrino flux}

The 1.44~MeV $pep$ $\nu_e$'s are an ideal probe to test the $P_{ee}$ transition region predicted by the MSW--LMA model because, thanks to the solar luminosity constraint,  its SSM predicted flux has a small uncertainty of 1.2\%~\cite{ssm_new}. The detection of $\nu_e$ from the CNO cycle would be the first direct evidence of the nuclear processes that are believed to fuel massive stars ($>$1.5 $M_{\odot}$). The CNO spectrum is the sum of 3 continuous spectra with end--point energies of 1.19 ($^{13}$N), 1.73 ($^{15}$O), and 1.74 MeV ($^{17}$F). The predicted CNO flux is strongly dependent on the solar modeling, being 40\% higher in the High Metallicity (GS98) than in the Low Metallicity (AGSS09) solar model~\cite{ssm,smp}. The detection of $pep$ and CNO $\nu_e$'s is more challenging than that of $^7$Be $\nu_e$'s, as their expected interaction rates are $\sim$10 times lower and because of the background in the 1-2 MeV energy range, the cosmogenic $\beta^+$-emitter $^{11}$C (lifetime: 29.4 min). Thanks to the extremely low levels of intrinsic background and to the novel background discrimination techniques Borexino  provided the first time measurement of the solar $pep$--$\nu_e$ rate and the strongest limit on the CNO solar $\nu_e$ flux to date \cite{pep}. 

The $^{11}$C background can be reduced by performing the Three--Fold--Coincidence (TFC) space--time veto after coincidences between muons and cosmogenic neutrons which are produced together with $^{11}$C in 95\% of the cases. The TFC veto relies on the reconstructed muon track and position of the neutron--capture $\gamma$-ray~\cite{mupa}. The rejection criteria were chosen to obtain the optimal compromise between $^{11}$C rejection and preservation of fiducial exposure. The resulting energy spectra before and after the TFC veto are shown in Figure~\ref{fig:pep}-Left. The residual $^{11}$C surviving the TFC veto is still a significant background. We exploited the pulse shape differences between $e^-$ and $e^+$ interactions in organic liquid scintillators~\cite{df} due to the finite lifetime of ortho--positronium as well as from the presence of annihilation $\gamma$--rays. An optimized pulse shape parameter was constructed using a boosted--decision--tree algorithm and used to discriminate $^{11}$C $\beta^+$ decays from  neutrino induced $e^-$ recoils and $\beta^-$ decays.  The analysis is based on a binned likelihood multivariate fit performed on the energy, pulse shape, and spatial distributions of selected scintillation events whose reconstructed position is within the fiducial volume~\cite{pep}. The fit included MC based distributions of the external $\gamma$--ray backgrounds. 

The best estimate for the $pep$--$\nu_e$ interaction rate in Borexino is (3.1 $\pm$ 0.6 (stat) $\pm$ 0.3 (syst)) counts/(day$\cdot$100ton)~\cite{pep}. The measured rate is to be compared to the predicted rate without $\nu_e$ oscillations, based on the SSM, of (4.47 $\pm$ 0.05) counts/(day$\cdot$100ton); the observed interaction rate disfavors this hypothesis at 97\% C.L. If this reduction in the apparent flux is due to $\nu_e$ oscillation to $\nu_\mu$ or $\nu_\tau$, we find $P_{ee} = 0.62 \pm 0.17$ at 1.44 MeV. Alternatively, by assuming MSW--LMA solar $\nu_e$ oscillations, this can be used to measure the $pep$ solar $\nu_e$ flux, corresponding to $\Phi_{pep}$ = (1.6 $\pm$ 0.3) $\times 10^8$ cm$^{-2}$ s$^{-1}$, in agreement with the SSM. Due to the similarity between the electron--recoil spectrum from CNO $\nu_e$'s and the spectral shape of $^{210}$Bi decay, whose rate is $\sim$10 times greater, we can only provide an upper limit on the CNO $\nu_e$ rate. Assuming MSW--LMA solar $\nu_e$ oscillations, the 95\% C.L. limit on the solar CNO $\nu_e$ flux is 7.7$\times 10^8$ cm$^{-2}$ s$^{-1}$. Our CNO limit is 1.5 times higher than the flux predicted by the High Metallicity SSM~\cite{ssm_new} and in agreement with both the high and low metallicity models.

\section{Conclusions}

The precision measurement of the $^7$Be solar $\nu_e$ rate validates the MSW--LMA solution of neutrino oscillation at 862~keV. Borexino measurement of the absence of the day--night asymmetry of the $^7$Be solar $\nu_e$ flux excludes the LOW solution at more than 8.5~$\sigma$ C.L. and, when combined with the other solar neutrino data, allows to select the LMA region of  neutrino oscillation parameters without assuming CPT invariance in the neutrino sector. The independent determination of the LMA solution obtained by Borexino with solar neutrinos only is reinforcing the consistency of our understanding of neutrino oscillations. Borexino has also achieved the necessary sensitivity to provide, for the first time, evidence of the rare signal from $pep$ solar $\nu_e$'s  and to place the strongest constraint on the CNO solar $\nu_e$ flux to date. This result raises the prospect for higher precision measurements of $pep$ and CNO interaction rates by Borexino, if the next dominant background, $^{210}$Bi, is further reduced by scintillator re--purification. Figure~\ref{fig:pep}-Right summarizes the current knowledge of the $P_{ee}$ for solar $\nu_e$'s to which Borexino results have contributed significantly. All data are in a very good agreement with the MSW--LMA solution of neutrino oscillations.

\section*{Acknowledgments}
The Borexino program is made possible by funding from INFN (Italy), NSF (USA), BMBF, DFG, and MPG (Germany), NRC Kurchatov Institute
(Russia), and MNiSW (Poland).  We acknowledge the generous support of the Laboratori Nazionali del Gran Sasso (LNGS).

\end{document}